# Ultrafast laser-driven topological phase patterning

*Jianyu Wu,[a] Arthur Niedermayr,[a] Gaolong Cao,[a] Oscar Grånäs,[b] and Jonas Weissenrieder*[,a]*

a. Light and Matter Physics, School of Engineering Sciences, KTH Royal Institute of Technology, SE-100 44 Stockholm, Sweden
b. Materials Theory, Department of Physics and Astronomy, Uppsala University, Uppsala, Sweden

*Corresponding author: jonas@kth.se

**Abstract**
Microscopic and dynamic control over quantum states is essential for bridging fundamental studies of material properties to device function. Realizing such control at combined high spatial resolution and ultrafast temporal precision remains a major challenge. Here, we demonstrate femtosecond laser-driven patterning of topological quantum states in the Weyl semimetal $WTe_2$. By engineering the excitation field into a transient optical grating, we spatially selectively and reversibly drive phase transitions between the topological Td and topologically trivial 1T* phases. Using ultrafast transmission electron microscopy, we directly visualize the formation of a periodic Td/1T* heterostructure, observe the propagation of a phase front, and analyze nanoscale confinement of coherently excited optical phonon modes. Our findings establish a platform for all-optical, spatially programmable, and reconfigurable control of quantum states, paving the way for optically addressable topological devices.

**Introduction**
Optical control of material properties with ultrashort laser pulses represents a rapidly evolving field of modern condensed matter physics.[1-3] By providing access to dynamics on femtosecond to attosecond timescales, analysis utilizing ultrashort laser pulses have unveiled a rich landscape of non-equilibrium phenomena—from light-field-driven electron tunneling to transient superconductivity—with promise for development of future applications in quantum communication and ultrafast information processing.[4, 5]

A particularly fertile ground for such control is found in quantum materials, where the interplay between electronic, spin, and lattice degrees of freedom can give rise to novel transient states.[6-8] A central challenge, however, lies in the spatially averaged nature of most ultrafast probes.[9, 10] Such probes obscures the spatial order—the microscopic arrangement of electronic and structural phases—that is fundamental to a material's function and potential for device integration.[11] Truly commanding the properties of these materials therefore requires ultrafast control with spatial specificity, targeting the relevant quantum degrees of freedom.

Structured optical fields present a powerful solution to this challenge. By sculpting light's intensity, phase, and polarization at the wavelength scale, one can selectively and coherently excite specific electronic, phononic, or spin subsystems.[12-14] This approach opens the door to creating and manipulating non-equilibrium quantum states with high spatial definition and ultrafast temporal precision.

In this work, we realize this potential by demonstrating ultrafast laser patterning of a topological quantum phase. Using a transient optical grating (TOG) generated by the interference of two femtosecond pulses, we drive a reversible structural transition between the topological Td and trivial 1T* phases in $WTe_2$. We directly visualize the resulting dynamics using ultrafast electron microscopy (UEM), achieving simultaneous nanometer and picosecond resolutions. This allows us to track the formation of a periodic Td/1T* heterostructure, observe the propagation of the phase front between the phases, and confirm the nanoscale confinement of a coherent optical phonon to the optically excited domains. Our findings establish a platform for the non-invasive, spatially programmable, and reconfigurable control of quantum states, paving the way for optically addressable topological devices.

**Topological phase patterning**

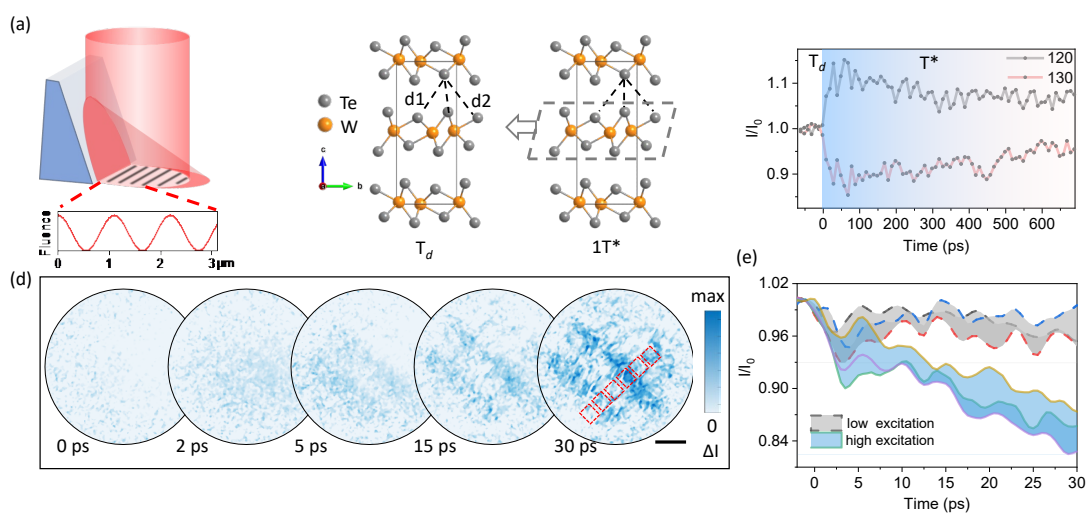

Figure 1. Light driven structural phase patterning in $WTe_2$. (a) Sketch of the TOG geometry formed by the interference between a direct laser beam and a beam reflected from a mirror placed in proximity to the sample. The resulting fluence variation in space is plotted in the lower panel. (b) Atomic structure models of the topological Td-$WTe_2$ phase with bond length d1 < d2 and the topologically trivial 1T* $WTe_2$ phase with d1 = d2.[7] (c) Time traces of the diffraction intensities of Bragg spots (120) and (130) during the structural transition between the Td and 1T* phases. (d) DF difference images at increasing time delay collected from the (130) spot from a $WTe_2$ sample excited by a TOG. A reference image, collected at -2 ps, is subtracted from each frame. Black scale bar: 1 μm. (e) Time dependent intensity traces from the six regions of interest (ROI) indicated in the last frame in (d). Three of the ROI:s are from spatial coordinates undergoing a phase transition (high local fluence), whereas other three are from regions where the transition does not take place (low local fluence).

At above threshold excitation, the ground-state Td-$WTe_2$ experiences a structural phase transition into a metastable 1T* phase, induced by relative shear displacements between neighboring layers along the crystallographic b axis.[7, 15] This interlayer shear modifies bond lengths (Figure 1b), redistributes the electron density, and changes the electronic topology. Td-WTe$_2$ is a Weyl semimetal with Weyl points that carry opposite chirality. The interlayer shear displacements drive the lattice towards a centrosymmetric structure, leading to annihilation of the Weyl points.[7, 16, 17] Thus, manipulation of the structural phase of WTe$_2$ by femtosecond laser provides a path for tuning topological invariants at high spatiotemporal resolutions. In diffraction experiments, the shear displacement manifests as alternating intensity changes in the Bragg spots aligned with the b-axis. Specifically, the (120) and (130) reflections show opposite intensity evolutions with increasing shift in b. Thus, these diffraction spots provide sensitive indicators for the transition (Figure 1c). The structural transition is completed within the first tens of ps after laser excitation and the metastable state persists for hundreds of ps before it recovers to the Td ground state.

The interlayer shear displacement of WTe$_2$ increases linearly with pump fluence,[15] which opens the possibility of manipulating the local structural phase and topology through spatially varying the laser fluence. We engineer the optical excitation into a TOG by splitting and recombining a femtosecond laser pulse at the sample plane. This results in a sinusoidal fluence modulation with a periodicity in our geometry of ~1.1 µm (Figure 1a). The TOG serves as a spatial template, imprinting alternating high- and low-fluence regions across the sample surface. Calculations for local pump fluence and periodicity are provided in Supporting information Figure S1.

The structural phase transition process in WTe$_2$ with TOG excitation is studied by dark field (DF) imaging. The DF images collected from the 130 Bragg spot show the formation of spatially patterned structural phases (Figure 1d). To enhance the dynamic image contrast, we subtract a reference image collected at -2 ps from all time delays (Figure S2). Dark fringes appear from ~2 ps time delay and become increasingly intense until 30 ps. The decrease in local Bragg intensity at the fringes indicates that the local lattice structure changes from Td toward 1T*. The fringes are fully established at 30 ps, with orientation and spatial periodicity in agreement with those of the TOG. Intensity traces from multiple regions of interest (ROIs) highlight the phase separated response (Figure 1e): ROIs at spatial sample positions excited with high local fluence exhibits a pronounced intensity drop (~20% at 30 ps), indicative of a local phase transition to the 1T* phase. The diffraction intensity from regions in between remains almost unperturbed. Superimposed on these transitions are coherent oscillations attributed to the optical $A_1$ shear phonon mode, which further reflect the lattice dynamics underlying the phase transition. These results demonstrate a powerful concept: by employing spatially structuring light, we can imprint nanoscale patterns of topological and trivial phases into a quantum material and track their evolution in real time.

## Phase propagation

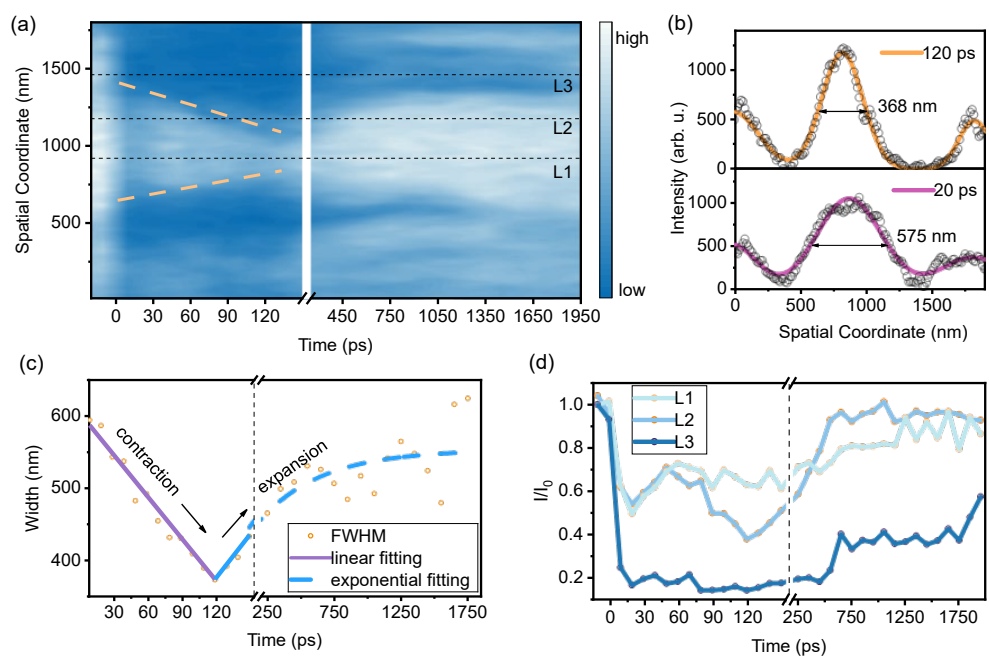

Figure 2. Phase evolution in space and time domains. (a) STCP generated from a DF image series using the (150) diffraction spot of WTe$_2$ excited with a TOG. Note: time steps are different before and after 150 ps. Two yellow dashed lines highlight interfaces between different domains. (b) Column intensity for the 20 and 120 ps time delays in the STCP. The solid lines represent Gaussian fits, with FWHM indicated for the central peak. (c) Extracted FWHM as a function of time delay. The results before and after 120 ps are fitted with two straight lines. The results after 150 ps are fitted with an exponential function. (d) Intensity profiles extracted from the black dashed horizontal lines in (a).

An understanding of the mechanisms governing phase separation and interaction between phases is critical for the local manipulation of topological invariants. Through DF imaging, we monitor the evolution of adjacent Td and 1T* phase domains and the position of the interfaces during the first 2 ns following laser excitation (Supplementary video 1). Sample regions subjected to local high-fluence exhibit an abrupt drop in intensity within 20 ps (300 – 700 nm and 1300 – 1700 nm, Figure 2a), indicative of transition into the 1T* phase. In contrast, low-fluence regions retain near-constant intensity, indicative of these regions remaining in the Td phase. Sharp boundaries form at the interface between these distinct responses. These boundaries are defined as the phase fronts. The DF contrast provides a direct visualization of the propagation of the phase-fronts.

Dark domains, corresponding to the 1T* phase, expand laterally around local fluence maxima, reaching maximum extension at ~120 ps before receding and eventually disappearing by ~2 ns. This temporal behavior reflects the excitation, growth, and relaxation of the metastable phase. A three-term Gaussian model was used to fit the column intensity for all time delays and extract the full width at half maximum (FWHM) of the central Td phase domain in the space time contour plot (STCP). Figure 2b shows extracted line profiles at 20 ps and 120 ps, where the decrease in domain size with time is evident. Notably, the phase-front propagates at a nearly constant velocity of ~1.0 nm/ps, as extracted from linear fits to domain widths at successive delays (Figure 2c). The extracted velocity is comparable to the shear-wave velocity in WTe$_2$ (~1.6 nm/ps),[18] suggesting that the phase propagation is strain-driven rather than thermally mediated, as thermal diffusion would be orders of magnitude slower (Figure S4).

Temporal DF analyses of locations within the TOG experiencing different local fluences provide insights into how the propagation of the phase front governs the dynamics of the local phase transition. The horizontal dashed lines in Figure 2a indicates three representative positions within the TOG: L1, located at the center of a Td phase domain (minimum local fluence, ~0.1 mJ/cm$^2$); L2, situated at the interface between the 1T* and Td phases (intermediate local fluence, ~1.8 mJ/cm$^2$); and L3, positioned at the center of a 1T* phase domain (maximum local fluence, ~3.5 mJ/cm$^2$). At L3, the intensity rapidly decreases to ~20% of its pre-excitation value (Figure 2d), corresponding to an interlayer shear displacement of ~23 pm along the b-direction. This displacement is notably larger than what has been observed for homogeneous excitation and closer to 37 pm required to complete the 1T* structural transition,[7, 15] reflecting the higher local excitation fluence achievable when energy is confined to nanometer-scale regions,[15] effectively mitigating thermal damage by minimizing heat accumulation. At the L2 position, the system experiences a two-step transition. The initial lattice distortion is insufficient to fully overcome the potential energy barrier and reach the 1T* local minimum. Because the system does not cross this barrier, a partial recovery toward the Td phase begins already before 30 ps. This recovery is subsequently interrupted and reversed at around 50 ps, leading to a secondary decrease in intensity that reaches a minimum at approximately 120 ps. This timescale coincides with the arrival of strain waves emanating from neighboring 1T* domains. In contrast, L1 exhibits only modest, transient lattice distortions (~12 pm), consistent with the relatively weak local optical excitation.

Taken together, these observations establish that the metastable 1T* phase propagates through lattice-mediated strain. Local laser excitation induces lattice distortions that generate a nonlocal release of stress into neighboring regions, thereby driving the propagation of the phase front. The comparable velocities of the phase front and the shear wave further support this interpretation. This mechanism illustrates how a spatially localized optical excitation can trigger collective structural rearrangements over micron-scale distances, providing a pathway for nonlocal phase engineering.

## Spatial and frequency mapping of optical phonons

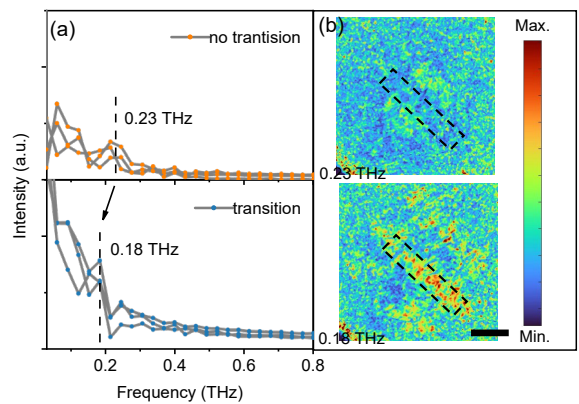

Figure 3. Mapping the excitation of optical phonons. (a) $A_1$ mode phonon softening with frequency indicated. The results are obtained by performing FFT on the data traces in Figure 1(e). (b) FFT amplitude mapping at two specific frequencies, 0.23 and 0.18 THz. The dashed rectangles highlight a region exhibiting phonon softening. The spatial FFT mapping of the $A_1$ shear mode is constructed from the DF image series in Figure 1. Each pixel represents the FFT amplitude ratio between the selected frequency and sum of total frequencies. Black scale bar: 1 μm.

Femtosecond optical excitation of $WTe_2$ not only drives structural phase transitions but also excites coherent interlayer shear phonons ($A_1$) that induce a lattice distortion analogous to that involved in the Td-to-1T* transition. The $A_1$ mode modulates the relative displacement of adjacent layers along the b-axis.[15] This periodic lattice displacement generates out-of-phase intensity oscillations of the (120) and (130) diffraction spots at a frequency close to 0.23 THz.[7] To spatially resolve the TOG excited phonon mode, we performed nano-beam diffraction (NBD) to probe structural oscillations at positions of different excitation fluence (Figure S8 and S9). The oscillating frequency exhibits a clear spatial dependence (Figure S10), confirming that the $A_1$ mode is modulated by the structured optical field. Notably, we observe a distinct phonon softening at a spatial periodicity matching that of the optically induced phase patterning. In regions experiencing the Td to 1T* transition, the $A_1$ phonon redshifts from 0.23 THz to approximately 0.18 THz (Figure 3a). These regions share the same periodicity and orientation as the TOG (Figure 3b) and display a complementary spatial distribution to the 0.23 THz regions, reflecting the fluence sensitivity of the excitation. Such frequency softening indicates a reduction in the interlayer forces as the lattice approaches the 1T* phase. In addition to structural effects, local lattice heating may also contribute to the observed frequency shift.[19, 20] These results demonstrate that optical fields can be engineered to confine coherent lattice vibrations to submicron domains.

## Multidimensional Mapping of Patterned States by U-STEM

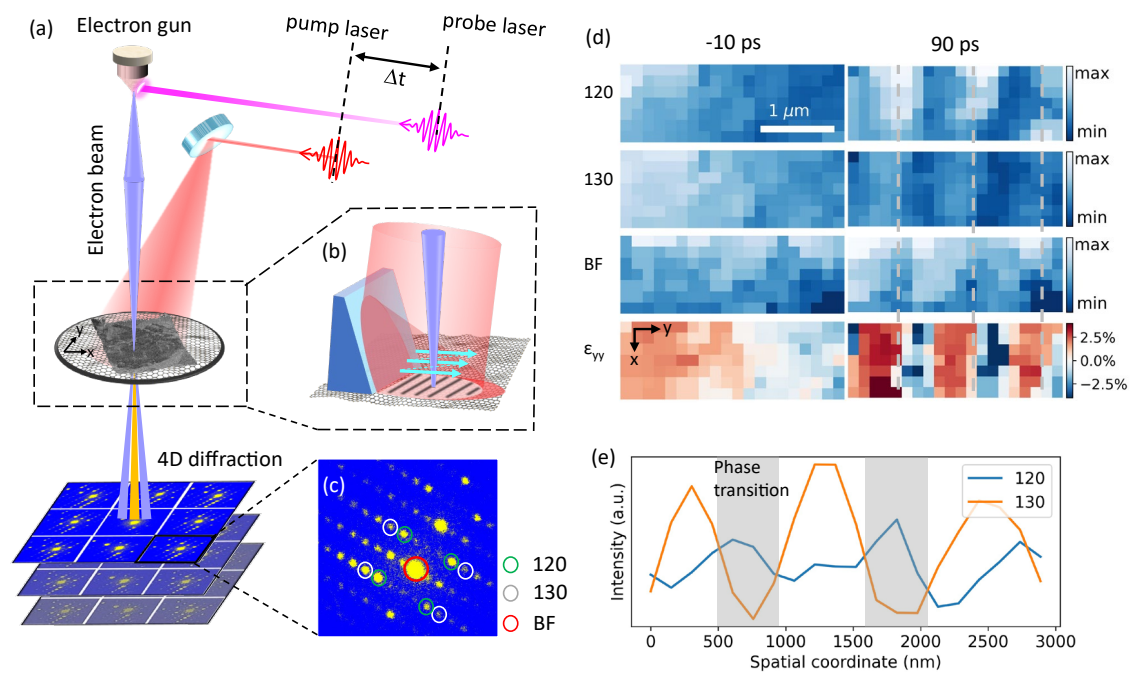

Figure 4. U-STEM analysis of local structural phase composition and strain. (a) Illustration of the U-STEM setup. Laser pulses are directed at the TEM cathode to generate electron pulses (probe). The probe position at the sample is controlled by the instrument scanning coils. A coupled laser pulse is directed to the sample (pump). A detector collects diffracted electrons as the probe scans the sample at different spatial and temporal coordinates. (b) Enlarged view of the formation of the TOG pump and the scanning path of the probe. (c) An electron diffraction pattern collected from $WTe_2$. Equivalent Bragg spots are summed up to construct virtual bright or dark field images. (d) Images constructed from diffraction spots (120), (130), transmitted beam (BF), and strain maps at -10 and 90 ps time delay, after a 2 × 2 median filter. Dashed lines indicate the position of local sample bending (in BF). (e) Intensity sum plot (smoothed) of DF after removal of the contribution from local structural bending.

Unambiguous identification of local structural transitions requires excluding contributions from other possible lattice distortions, such as photoexcited coherent acoustic phonons. This requirement necessitates to move beyond the single-lattice-plane information accessible in conventional DF or bright field (BF) imaging. Four-dimensional scanning transmission electron microscopy (4D-STEM) overcomes this limitation by recording diffraction patterns from multiple lattice planes at each probe position, thereby capturing a comprehensive description of local crystallography and strain fields.[21] Here, we extend 4D-STEM into the time domain by integrating femtosecond optical excitation with stroboscopic electron probing, realizing picosecond-resolved ultrafast 4D-STEM (U-STEM, Figure 4a) of the phase-patterned $WTe_2$ system. A ~200 nm electron probe was rastered across the sample, recording diffraction patterns at each spatial coordinate and pump–probe delay. DF and BF images can be reconstructed by integrating intensity from selected Bragg reflections, allowing visualization of structural states from multiple lattice planes (Figure 4d). These reconstructed images exhibit contrast fringes whose periodicity and orientation match the exciting TOG, but with slight position shift between frames, arising from the combined effects of local structural bending of the sample and phase transition. The contrast in the BF channel reflects periodic structural sample bending, similar to previous observations of acoustic phonon–driven deformation,[22] while the DF images reveal the superimposed phase-transition pattern. The local phase distribution can be isolated by subtraction of the bending contribution extracted from the BF signal (Figure S7). The validity of this method is evidenced by the anti-phase intensity distribution between 120 and 130 (Figure 4e). Additionally, U-STEM provides direct access to time-resolved strain fields by tracking reciprocal-space shifts of Bragg peaks. The $\varepsilon_{yy}$ strain component display a striped profile aligned with the optical excitation. Since the lattice parameter remains constant during the Td→1T* transition, the detected strain arises predominantly from photoinduced sample bending. These results demonstrate the capability of U-STEM to simultaneously probe multiple diffraction channels, providing a multidimensional view of excited-state dynamics. A single line scan dataset resolves both the phase patterning and coherent optical phonons (Figure S6). Such rich diffraction dataset allows quantitative mapping of structure, crystal orientation, strain and even electric and magnetic fields across space and time,[23, 24] enabling U-STEM as a versatile platform for probing ultrafast, spatially resolved phenomena with nanometer spatial and picosecond temporal resolution.

**Estimating Topological Robustness in finite domains**
The topological classification is rooted in global properties of the electronic band structure. A key question for optically patterned $WTe_2$ is whether finite Td domains embedded in a 1T* matrix preserves their topological character. In Weyl semimetals, finite-size confinement can cause surface states from opposite sides to overlap and hybridize, opening gaps that eliminate Weyl crossings. In the case of a type-II Weyl semimetal, we must ensure that finite-size effects do not remove protected band-crossings. Robust topology therefore remains well-defined when the domain size is much larger than the decay length of localized Wannier states.

We investigate this by parametrizing a tight-binding model, constructing finite slabs and investigating the hybridization gap $\Delta E$. Further, following the framework developed by Benito-Matias *et al.*[25] for finite sized slab geometries, we derive a continuum expression for the hybridization gap that is used to estimate the hybridization gap for in-plane domains. (see S.I. for details) The parameters of the model $\xi$ are estimated by fitting to the parametrized tight-binding model for finite geometries. From an experimental perspective, also angular resolved photoemission may be used.[26]

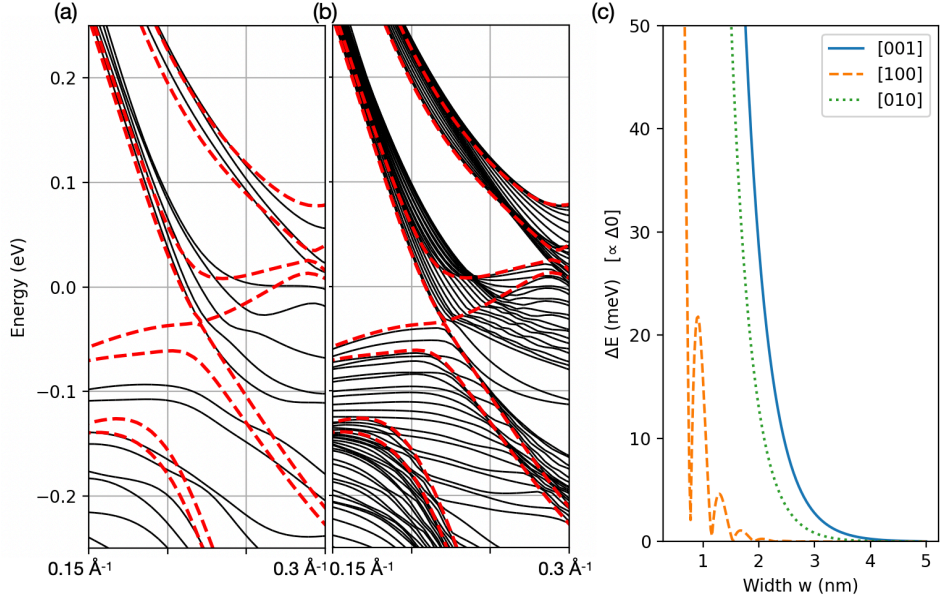

Figure 5. Band structure calculation. (a) shows the band-structure of the bulk (dashed red lines) around one of the Weyl-nodes, superimposed on the band-structure for a finite 2-layer slab in the [001] direction (black), (b) shows the bulk bandstructure superimposed on the 10-slab finite bandstructure. (c) shows the continuum model, with parameters fitted so three finite calculations (number of layers being 2, 3 and 5, corresponding to 2.8, 4.2 and 7 nm).

The Weyl-points in Td WTe$_2$ are located in the $k_x k_y$ plane. The coordinates in cartesian coordinates as well as the Chern numbers are listed in table S2, matching well to published results.[16] As an upper bound to the decay length of the surface/interface state, we investigate confinement in the z-direction (as imposed by the sample thickness). In Figure 5a and b, we show the bulk band structure along a path in reciprocal space, parallel to the $\Gamma - X$ direction, but offset in the y-direction to ensure that the Weyl-node (node 4 in table S2) lies on the path. Panel (a) shows a two unit-cell (2.8 nm) cell, indicating a substantial gap and no Weyl node beyond the thermal energy scale of ~25 meV (corresponding to room temperature). At 10-layer thickness (14 nm), the crossing has acquired bulk-like properties. Figure 5c shows the result of the continuum model, for three different directions. Since experimental domains are hundreds of nanometers wide and tens of nanometers thick, the hybridization gap can be considered zero for all widths $w$ regardless of orientation of domain or cut.

**Outlook**

Optical fields offer a transient and programmable route to define topology with nanometer spatial and picosecond temporal resolutions. The interference patterns demonstrated here can be shaped into more complex geometries using multiple beams, enabling deterministic control over size, symmetry, and dimensionality.[22] Pattern periodicity can be tuned via the wavelength and interference angles, reaching nanometer-scale precision.[13] The resulting spatially patterned structural phases and topological states enable tailored distributions of material properties, opening avenues for advanced optoelectronic and quantum devices.[27] For instance, grating-patterned crystallographic phases modulate the nonlinear susceptibility in a sinusoidal distribution. Second-harmonic generation from such transient grating symmetries enables tunable emission angles that directly depend on the grating periodicity (Figure S5).

**Conclusions**
We have demonstrated transient phase patterning of the topological material Td-WTe$_2$ and spatially selective excitation and detection of optical phonons. Using transient optical gratings, we drive a structural phase transition in nanoscale regions of WTe$_2$, resulting in a Td – 1T* phase heterostructure. The phase front at the interface between the 1T* and Td domains propagates at a velocity comparable to that of the shear strain wave in the material, revealing a nonlocal phase excitation mediated by the interplay between electronic excitation and lattice strain. Furthermore, the A$_1$ optical phonon mode is confined to nanoscale regions aligned with the optical grating, demonstrating precise spatial control of coherent lattice dynamics. We demonstrate that red shifts in the local phonon frequency can be resolved spatially. Tight-binding simulations confirm the robustness of the nano-patterned topological structure. U-STEM provides multidimensional observations of the transient structural states and can be used to pinpoint the spatial locations of the two different phases. These findings establish a novel platform for manipulating topological states in 2D materials with ultrafast temporal control and nanoscale precision, opening avenues for direct laser-written circuits and optically controlled quantum devices.

**Method**
**Sample Preparation**
WTe$_2$ samples were prepared by mechanical exfoliation using adhesive tape[28] from a bulk single crystal of Td-WTe$_2$ (2D Semiconductor, USA). The sample was then transferred to a piece of polydimethylsiloxane (PDMS) and checked under an optical microscope. Desired flakes were aligned and attached onto a 50 nm Si$_3$N$_4$ TEM grid (Ted Pella) under the optical microscope. A slanted surface of a Si TEM grid is used as a mirror to partially reflect the pump laser beam. The grid is placed on top of the selected sample flake, with its slanted surface oriented towards the pump laser beam. The grid was coated with aluminum and exhibited a reflectivity of approximately 50%. The laser is aligned such that a portion of the beam is reflected off the slanted surface, while the remainder continues along its original path. The reflected and direct beams interfere at the sample plane, generating a TOG. The resulting laser fluence distribution is governed by the amplitude and phase difference between the interfering beams.

**Ultrafast electron microscope (UEM) setup**

The time-resolved experiment was performed in an UEM (based on a JEOL JEM 2100) operating at 200 kV. The electrons are detected by a hybrid pixel detector (CheeTah T3, Amsterdam Scientific Instruments). The electron bunches (probe) were generated through photoemission from a guard ring $LaB_6$ cathode by 258 nm laser pulse illumination. The sample was excited by a λ = 1030 nm and 300 fs laser pulse (pump) (Tangerine, Amplitude Systemes). The laser was focused to a spot with FWHM of ~200 µm at the sample plane. The time delays between the pump and probe pulses were controlled using a motorized delay stage varying the optical path length of the pump beam. The $WTe_2$ samples were excited at a repetition rate of 12 kHz. Detailed information of the instrumental setup can be found in ref [29]. The experiments were performed at room temperature. In dark field imaging, an objective aperture was inserted to selectively capture the diffraction spot (130) or (150) while blocking all other spots. Consequently, the collected DF images include dynamic information only from the lattice plane corresponding to the selected diffraction spot.

**U-STEM**

The UEM is set to operate in scanning mode by directing a nearly parallel nano-electron beam across a sample region. A 10 µm diameter condenser aperture is used to reduce the electron beam size. The spatial position of the beam is controlled by scanning coils. After scanning the designated sample region, the delay stage shifts to the next time delay. At each scanning step, the direct electron detector captures the resulting diffraction patterns. In line scanning, the scanning trace consists of 40 pixels in one dimension with a step size of 116 nm. In 2D scanning, the scanning area is a 20*8 pixels region, with a step size of 152 nm.

**Theory**

The Weyl-points and the associated coordinates listed in table S2 are calculated using WannierTools [30]. The underlying Wannier functions were produced with Wannier90 [31] based on the Vienna Ab Initio Simulation package (VASP) [32, 33] using input and structural parameters from [34].

**Supplementary Material**

Supplementary information including fluence calculation with transient optical grating, additional sample images, COMSOL model and simulation results, U-STEM data processing method, characterization of the electron beam in NBD mode, topological robustness calculation are available.

**Conflict of interest**

The authors have no conflicts to disclose.

**Acknowledgements**
Jianyu Wu gratefully acknowledges the Chinese Scholarship foundation (CSC) for a doctoral fellowship. A.N. acknowledges funding from the Swiss National Science Foundation (SNSF) through project P500PT_214469. This research was made possible through the generous funding of the Knut and Alice Wallenberg Foundation (2012.0321 and 2018.0104), the Swedish Research Council (VR 2021-04379), the ARTEMI national infrastructure through VR grant 2021-00171 and Strategic Research Council (SSF) grant RIF21-0026. OG acknowledges funding from the European Research Council through the Synergy Grant 854843 - FASTCORR. The computations were enabled by resources provided by the National Academic Infrastructure for Supercomputing in Sweden (NAISS) and the Swedish National Infrastructure for Computing (SNIC) at NSC and PDC and Uppmax partially funded by the Swedish Research Council through grant agreements No. 2018-05973. and No. 2022-06725 and No. 201805973.

**Supporting information to:**

# Ultrafast laser-driven topological phase patterning
*Jianyu Wu,[a] Arthur Niedermayr,[a] Gaolong Cao,[a] Oscar Grånäs,[b] and Jonas Weissenrieder*,[a]*

a. Light and Matter Physics, School of Engineering Sciences, KTH Royal Institute of Technology, SE-100 44 Stockholm, Sweden
b. Materials Theory, Department of Physics and Astronomy, Uppsala University, Uppsala, Sweden

**Transient optical grating**

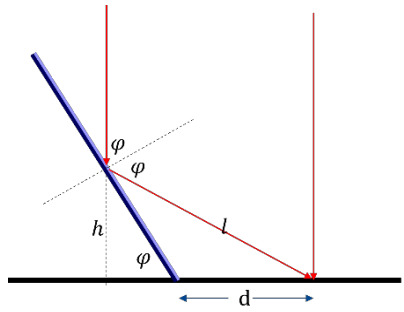

Figure S1. Calculation of local fluence with transient optical grating.

Careful alignment of the pump laser beam to the intersection of the edge of a mirror placed in close proximity to the sample splits the laser beam into two paths. The to the sample direct laser beam and the beam reflected from the slanted side face of the mirror will interfere at the sample plane. The condition for constructive interference determines the spatial period between the interference fringes of the transient optical grating. The incident laser beams are marked as red arrows. The reflecting slanted side face is marked in blue, and the sample plane is marked in black. By applying the conditions for constructive interference and phase shift upon reflection we can formulate an expression for the phase difference Δϕ between the direct and reflected beams from the path difference Δd:

$$\Delta\varphi = \frac{\Delta d}{\lambda} * 2\pi = (l-h) * \frac{2\pi}{\lambda} = \frac{2\pi d}{\lambda} * \frac{sin\varphi(1+cos2\varphi)}{cos\varphi}$$

Where λ is the wavelength of incident laser, ϕ is the angle between the mirror surface and the sample plane (55° in our geometry). The local fluence with two interfering beams is calculated from:

$$F(r) = F1 + F2 + 2\sqrt{F1F2}\cos(\Delta\varphi(r))$$

Where F1 and F2 are the incident fluence of the individual beams. $\Delta\phi(r)$ is the phase difference between the two beams at position $r$ (distance from the edge of the mirror). The fluence of the direct incident laser (F1) is 5 mJ/cm², while the reflected laser (F2) is:

$$F2 = F1 * R$$

Where R is the reflectivity of the mirror (here R = 0.5).

**Characterization of sample**

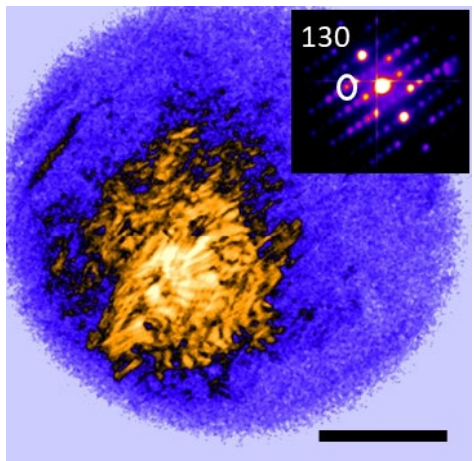

Figure S2. Dark field image at -2 ps with a diffraction pattern as inset.

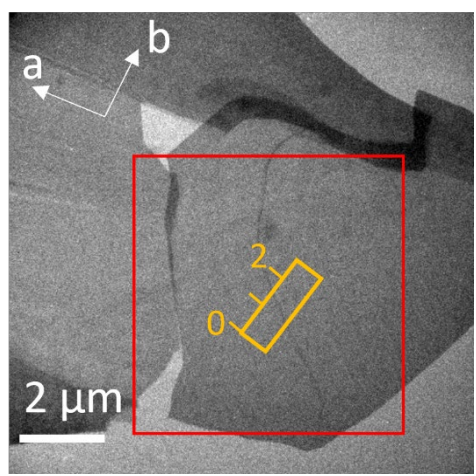

Figure S3. TEM image of the sample in Figure 2 of the main text. The yellow box indicates the region for space-time contour plot in Figure 2(a).

**COMSOL simulation**

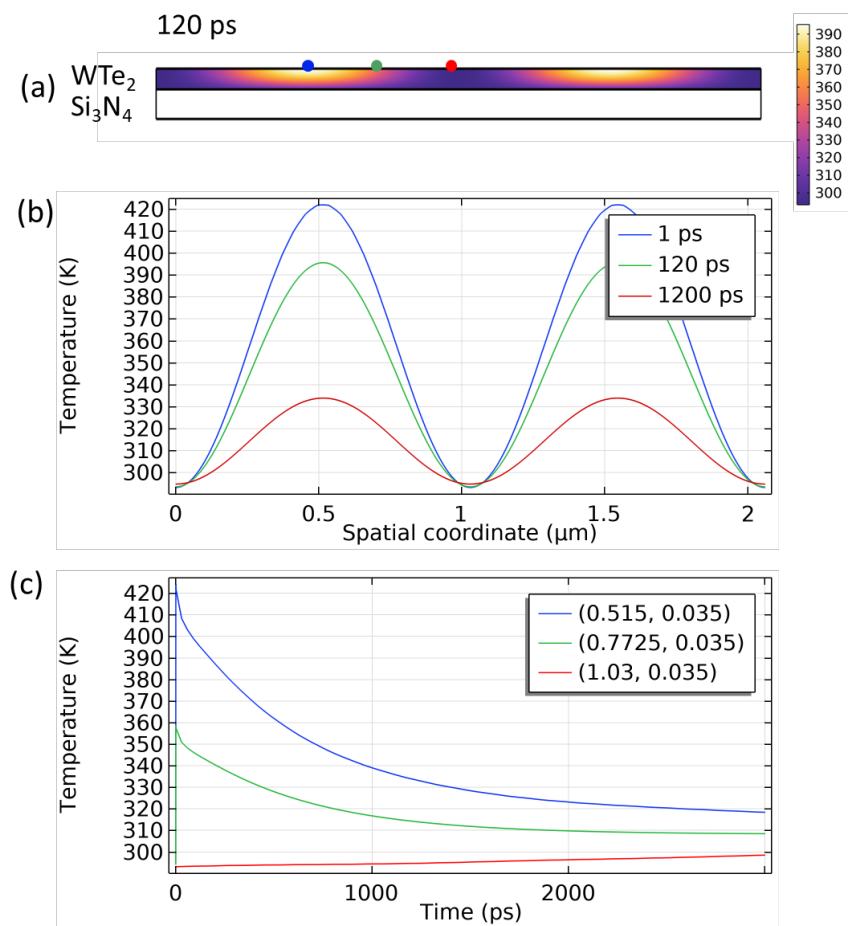

Figure S4. (a) Model construction in COMSOL Multiphysics. The panel shows the simulated temperature distribution at 120 ps time delay after excitation with a TOG. (b) Temperature profile along the top surface of the model at three different time delays. (c) Simulated temporal temperature evolution at the three positions indicated by filled circles in (a) in corresponding colors.

COMSOL Multiphysics is employed to simulate the temperature evolution of a $WTe_2$ sample under transient optical grating excitation. The Heat transfer in solid module was utilized for the simulations. The relevant material specific parameters for $WTe_2$ used in the simulation can be found in Table S1. The model comprises a 35 nm $WTe_2$ flake and a 50 nm $Si_3N_4$ substrate with the interface located at z = 0 nm as shown in Figure S4(a). The simulation cell boundaries are set in periodic conditions as for an infinite sample size. The top and bottom boundaries of the sample are left free. The laser induced lattice heating is simulated by applying a transient heat source, which represents the energy transferred, per unit volume, from the laser pump to the material. The temporal profile of the heat source adheres to the temporal pulse length of the pump laser and is modeled as a 300 fs pulse applied at time 0. The local power of the heat source is modeled with a gradient in the thickness direction, to simulate local light absorption modeled by the Beer–Lambert law in the thickness direction. The power of the heat source follows a sinusoidal function along the horizontal axis (in-plane of the sample) of the model.

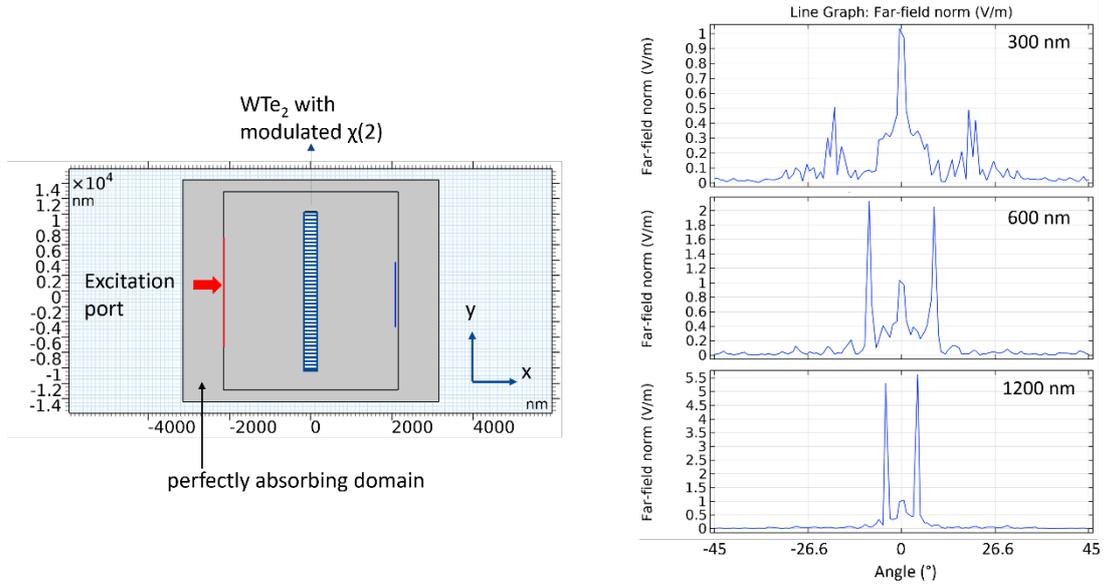

Figure S5. Second harmonic generation simulated by FEM. Left: constructed COMSOL model for WTe$_2$ with TOG grating modulated susceptibility. Right: far-field radiation of second harmonics for different grating periodicities.

The model comprises a WTe$_2$ sample of 30 nm thickness (x axis). The nonlinear susceptibility $\chi^{(2)}$ is modulated in gratings (sinusoidal function between Td and 1T*) in the y direction of the model. For Td phase, $\chi_{Td}^{(2)} \approx 95$ pm/V.[1] For 1T* phase, $\chi_{1T*}^{(2)} = 0$. The incident wave is polarized in the y-direction and launched at the fundamental frequency $f1$ (291 THz) with power at $1 \times 10^7$ W/m using an excitation port. The second harmonic generation traces (right panel) are for a distance of 2 μm after sample. The boundary of the simulation cell is modeled as a perfectly absorbing layer to represent an open and nonreflecting infinite volume. The Electromagnetic Waves, Frequency Domain interfaces are defined for the fundamental frequency $f1$ and the second harmonic frequency $2*f1$. The two interfaces are coupled using a polarization feature added to each of the interfaces.

For the fundamental interface, the polarization is given by:

$$P_{1y} = 2dE_{2y}E^*_{2y}$$

and for the second harmonic interface the polarization is given by

$$P_{2y} = dE^2_{1y}$$

where d is a nonlinear coefficient and calculated by $d=\varepsilon_0*\chi^{(2)}$, $E_{1y}$ and $E_{2y}$ are the y-component of the electric field at the fundamental frequency and the second harmonic generation. When the spatial periodicity of the grating increases from 300 nm to 1200 nm, the second harmonic far-field radiation shows increasingly sharp lines at the first order interference fringes.

**U-STEM**

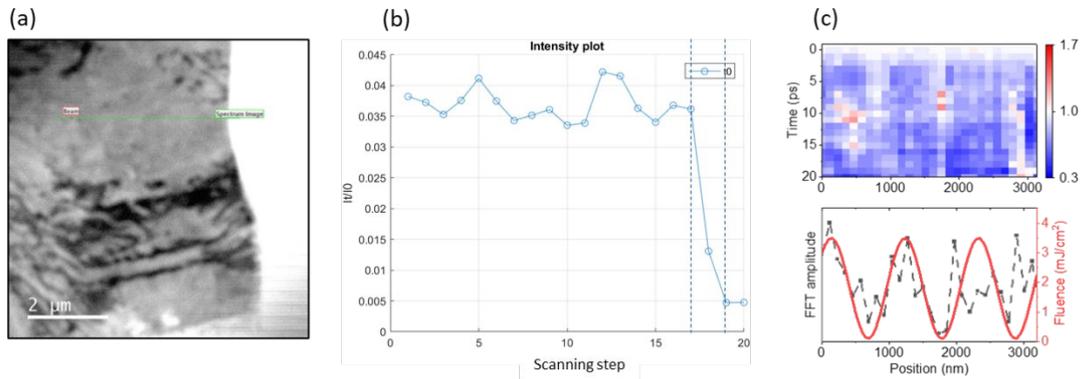

Figure S6. U-STEM line scan across a sample edge. (a) STEM image showing the scanned region (green line). (b) Intensity profile of the summed (130) diffraction spot as a function of scanning steps. (c) Upper panel: Dark-field image reconstructed from the (130) reflection. Lower panel: Calculated local pump fluence (red line) and the FFT amplitude at 0.23 THz (dashed line), extracted from the time-domain signal at each spatial column in the upper panel.

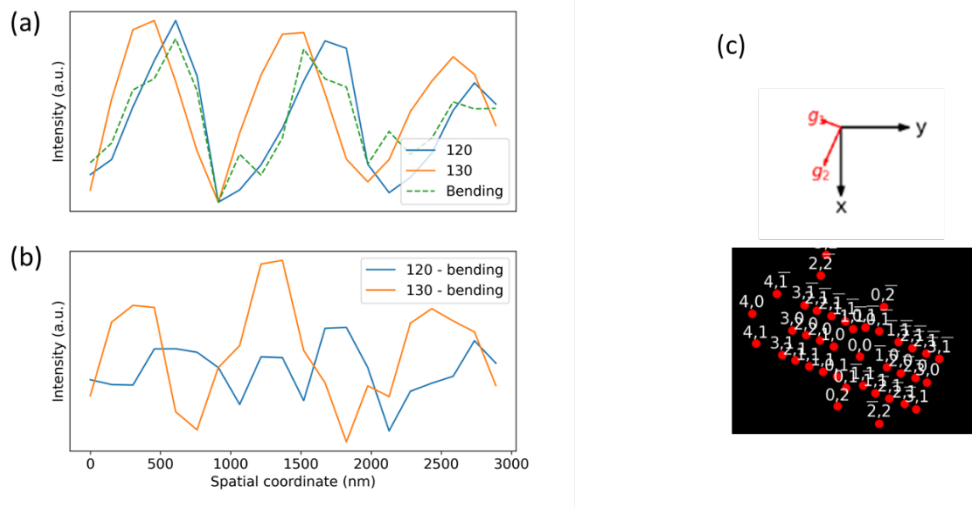

Figure S7. (a) Normalized intensity of the 120 and 130 spots and BF (Bending) from the dataset in Figure 4d of the main text. (b) Intensity traces for the 120 and 130 Bragg spots after removing the contribution from sample bending (BF in a). (c) Bragg spots used for strain calculation, with basis vector g1 and g2 indicated.

The probe beam in U-STEM mode was scanned across the sample edge with a step size of 132 nm. From the 2 steps abrupt intensity drop, the effective probe diameter is determined to be around 200 nm. The images and strain map were calculated using Py4dstem.[2] For the reconstructed BF and DF images, intensities were summed along the vertical direction. The BF reflects sample bending and is insensitive to the structural phase transition. The contribution from local sample curvature was therefore obtained by subtracting the BF intensity from its maximum value. This bending component was then removed from the (120) or (130) DF signals, allowing for separating of the structural phase-transition contrast from geometric bending effects.

**Nano beam diffraction**

In nano beam diffraction mode (NBD), the electron beam was converged to a spot with a diameter of 330 nm. The characterization of the electron beam is shown in Figure S7. By scanning the beam across the sample, we find a location for minimal temporal oscillation of the intensity of spot (130) at position 1. This implies a region of low local pump fluence. By shifting the probe beam 330 nm perpendicular to the direction of the TOG fringes, two additional probe positions (2 and 3) are selected to illustrate the evolution of lattice dynamics. The sample position and collected diffraction intensity trace are shown in Figure S8 and Figure S9.

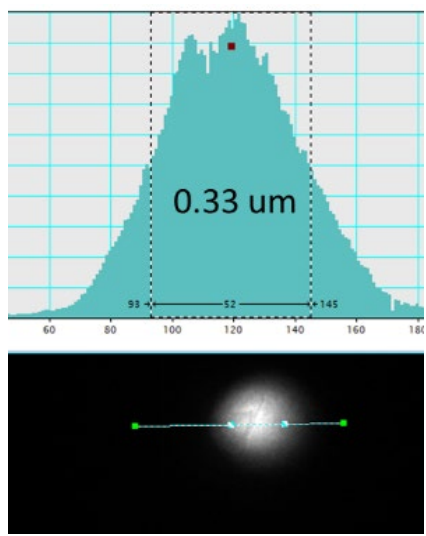

Figure S8. Characterization of the electron beam in NBD mode. The full width at half-maximum of the beam is 330 nm.

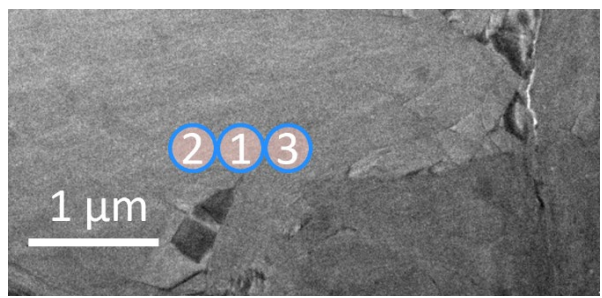

Figure S9. Sample image and three probe positions in NBD mode.

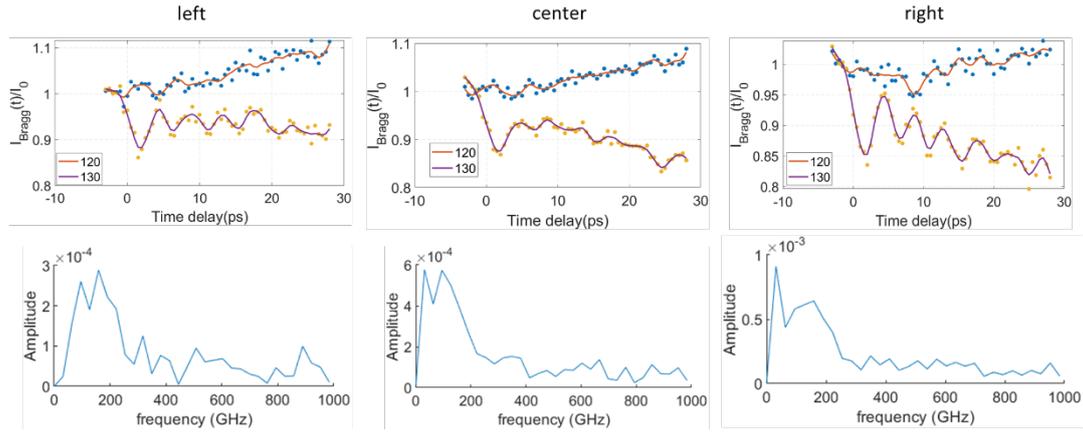

Figure S10. Upper panels: Intensity traces for diffraction spots 120 and 130 at different probe positions. Lower panels: FFT spectrum of spot 130. The results are collected from positions 1 (center), 2 (left) and 3 (right).

Table S1. Material properties for WTe$_2$ used in simulations.

| Property | Value | Unit | Reference |
| --- | --- | --- | --- |
| Density | 9863 | kg/m³ | [3] |
| Thermal conductivity | k11=1.7, k12=1.6, k13=1.0 | W/(m·K) | [4] |
| Heat capacity at constant pressure | 170 | J/(kg·K) | [5] |
| Absorption coefficient (at 1030 nm) | 2.59E7 | 1/m | [6] |

**Topological Robustness calculation**

Table S2. The coordinates in cartesian coordinates as well as the Chern numbers from tight-binding calculations.

| ID | Kx | ky | kz | C |
|---|---|---|---|---|
| 1 | -0.2195 | -0.0451 | 0.0 | +1 |
| 2 | -0.2202 | -0.0389 | 0.0 | +1 |
| 3 | 0.2195 | 0.0451 | 0.0 | +1 |
| 4 | 0.2202 | 0.0389 | 0.0 | +1 |
| 5 | -0.2194 | 0.0456 | 0.0 | -1 |
| 6 | -0.2202 | 0.0387 | 0.0 | -1 |
| 7 | 0.2194 | -0.0456 | 0.0 | -1 |
| 8 | 0.2202 | -0.0387 | 0.0 | -1 |

*Derivation of continuum model for hybridization gap*

Consider a slab of a Weyl semimetal of thickness $w$ with surfaces at $z = \pm w/2$. The starting point is the low-energy Hamiltonian used in Ref. [7] [ Eq.(1)]:

$$H_\chi(\mathbf{k}) = \epsilon_0(\mathbf{k})\,\mathbb{1} + M(\mathbf{k})\sigma_z + v_x \chi \sigma_x k_x + v_y \chi \sigma_y + v_z \sigma_z (k_z - \chi k_0)$$

with $\chi = \pm 1$ the node chirality. Here $\epsilon_0$ and $M$ are even in $\mathbf{k}$, including tilt terms for type-II WSMs. In Ref. [7], a specific $c_i, m_i$ parametrization of Eq.(1) is used, our derivation holds for the general anisotropic, tilted form.

Based on this Hamiltonian, we now derive a compact formula for the finite-thickness hybridization gap $\Delta E(\mathbf{k}_\parallel)$ of two counterface surface states, as developing at an interface, including its exponential decay with $w$ and the tilt-induced oscillatory envelope. For a single surface, we seek states decaying along the slab normal, using $k_z \to -i\partial_z$. Following the ansatz of Eq.(4) in Ref. [7]:

$$\psi_{\mathbf{k}_\parallel z} = \sum_j A_j\, e^{-\lambda_j z} \eta_j(\lambda_j)$$

where the decay parameters $\lambda_j$ are obtained by substituting into the eigenproblem defined in Eqs.(5)-(6) in Ref. [7]. For type-II WSMs, the roots $\lambda$ generically have nonzero real and imaginary parts, giving oscillatory leakage into the bulk.

Under the assumption that we have a dominant decay constant $\kappa(\mathbf{k}_\parallel) = \min_j Re[\lambda_j]$ and oscillation wave number $\alpha(\mathbf{k}_\parallel) = Im[\lambda_j]$, they are read off from the semi-infinite solution. With two surfaces a distance $w$ apart, the general finite-slab boundary problem in Ref. [7] (Eqs. (12)-(14) or equivalently Eqs.(A23)-(A25)) yields the exact dispersion $E_\pm(\bm{k}_\parallel)$ in terms of $\lambda_{1,2}$.

Near a crossing of the two isolated-edge dispersions, degenerate perturbation theory in the top/bottom subspace defined by $\{\psi_T, \psi_B\}$ subspace gives

$$E_\pm \approx \frac{E_T + E_B}{2} \pm \frac{1}{2}\sqrt{(E_T - E_B)^2 + 4|t_\perp|^2}$$

Where $t_\perp(\bm{k}_\parallel) = \langle \psi_T | H | \psi_B \rangle$.

From the explicit slab solutions, the two surface modes have envelopes

$\psi_T(z) \propto e^{-\kappa(z-\frac{w}{2})} e^{i\alpha(z-\frac{w}{2})}$ and $\psi_B(z) \propto e^{+\kappa(z+\frac{w}{2})} e^{i\alpha'(z+\frac{w}{2})}$ in the interior.

Sandwiching $H$ between them, the $z$-integral gives $t_\perp \propto \dfrac{(\eta_T^* v \eta_B) e^{-\kappa w} \sin\left(\frac{\Delta\alpha w}{2}\right)}{\frac{\Delta\alpha}{2}}$, where $v$ is the local $z$-coupling operator (e.g. $v_z \sigma_z(-i\partial_z)$) $\Delta\alpha = \alpha - \alpha'$.

This reproduces the exponential gap scaling and oscillatory envelope seen in the analytic gap estimate for type-B states [Eq.(29) in Ref. [7]]. Combining the perturbative splitting with the overlap form of $t_\perp$, we obtain

$$\Delta E(\bm{k}_\parallel) \approx 2\Delta_0(\bm{k}_\parallel) e^{-\kappa(\bm{k}_\parallel)w} \left|\text{sinc}\left(\frac{\Delta\alpha(\bm{k}_\parallel)w}{2}\right)\right|.$$

Where $\kappa$ and $\alpha$ are from the semi-infinite solution [Eqs. (A14)-(A17) in Ref. [7]], and $\Delta_0$ encodes the short-range spinor coupling. In the limit of large width, $\lambda_{1,2}$ reduces to their isolated-edge values found in [Eqs. (A14)-(A15) in Ref. [7]], and $\Delta E$ matches the perturbative $2|t_\perp|$ form. We now proceed to fit $\Delta_0(\bm{k}_\parallel)$, $\kappa(\bm{k}_\parallel)$, and $\Delta\alpha(\bm{k}_\parallel)$ to explicit calculations based on WannierTools, using thicknesses of 2, 3, and 5 unit cells, equivalent to 2.8, 4.2 and 7 nm thicknesses. Note that the minimum of the hybridization gap is not at a constant $\bm{k}$, we use the value of the original Weyl-point of the bulk. This rough estimate is motivated by the fact that the gap closes to below the thermal limit of ~25 meV (corresponding to room temperature) even in our explicit calculations for finite slabs of a few nm width. For directions other than $z$, $\bm{k}_\parallel$ is calculated based on the Weyl-node position, but the short range decay $\Delta_0$ is used based on the assumption that the decay is slowest for states terminating at vacuum (as opposed to the comparatively smooth matching conditions at phase boundaries).